# Path Planning and Robust Path Tracking Control of an Automated Parallel Parking Maneuver

Xincheng Cao, Levent Guvenc

Automated Driving Lab, Ohio State University

## Abstract

Driver's license examinations require the driver to perform either a parallel parking or a similar maneuver as part of the on-road evaluation of the driver's skills. Self-driving vehicles that are allowed to operate on public roads without a driver should also be able to perform such tasks successfully. With this motivation, the S-shaped maneuverability test of the Ohio driver's license examination is chosen here for automatic execution by a self-driving vehicle with drive-by-wire capability and longitudinal and lateral controls. The Ohio maneuverability test requires the driver to start within an area enclosed by four pylons and the driver is asked to go to the left of the fifth pylon directly in front of the vehicle in a smooth and continuous manner while ending in a parallel direction to the initial one. The driver is then asked to go backwards to the starting location of the vehicle without stopping the vehicle or hitting the pylons. As a self-driving vehicle should do a much better job repeatably than a driver, a high order polynomial path model is built along with speed profiling to start and stop smoothly at the ends of the path without large longitudinal and lateral accelerations. In contrast to the long horizon, higher speed path planning and path tracking control applications in the literature, this paper treats low speed and very short horizon path planning and path tracking control with stopping and direction reversal. The path is constructed using a segmented polynomial fit optimization routine that guarantees path curvature smoothness. A linear path-tracking model is utilized as the basis of the designed control system consisting of a disturbance observer based curvature rejection filter and a speed-scheduled, parameter-space robust PID controller. Simulation studies are conducted to analyze the tracking performance of the combined control system, and results indicate that it has better performance compared to other common control systems such as standalone PID controller and combined PID and feedforward control.

## Introduction

Vehicle automation has seen rapid advancements in recent years, as Advanced Driver Assistance Systems (ADAS) are becoming increasingly more common [1]–[5]. One of the goals of self-driving vehicles is to achieve better-than-human performance. Such objective means that the automated driving system should be able to handle the human driver license test in a satisfactory manner. This becomes the motivation behind the analysis performed in this paper.

A self-driving vehicle will first need to plan the right path for the driving exam parking emulation maneuver as path planning is an important component for path following operations. There have been multiple methods for generating collision-free and dynamically feasible paths. Some examples include elastic band method as discussed in [6], potential field method first illustrated in [7], A* algorithm first demonstrated in [8] and its further development such as the D* algorithm discussed in [9]. Additionally, [10] provides some additional path description methods such as Bezier curves, clothoids and polynomial spline.

Path-tracking control is another critical component for satisfactory path following performance. There is a range of various control methods available for path-tracking problem, including but not limited to model predictive control as described in [10], pure-pursuit controller as discussed in [11], [12] and Stanley controller as illustrated in [13]. It turns out that disturbance observer (DOB), or also recently called the curvature rejection filter [10] as well, is very suitable for path-tracking control. [14] provides detailed description of the disturbance observer method in its application to on road vehicle yaw stability control. Parameter-space designed multi-objective PID control is also a good general purpose robust control method. References [15] and [16] describe this method in detail.

The organization of the rest of the paper is as follows. The maneuverability test problem statement and polynomial spline path synthesis are treated next, followed by the introduction of the linear path-tracking model. The speed and preview distance scheduling routine is presented next. This is followed by disturbance observer and speed-scheduled parameter-space PID controller design. A simulation study is used to demonstrate the efficacy of the method proposed here. The paper ends with conclusions and recommendations.

## Maneuverability Test and Path Generation

This section aims to present the segmented polynomial fit optimization procedure to generate a reference path for the control system. This approach is also discussed in [10].

The schematic for the maneuverability test setup is displayed in Figure 1. The vehicle starts from a standstill and begins the maneuver by going straight and forward, before picking either one of the two directions (illustrated in Figure 1(a) and Figure 1(b) respectively) for a single lane-change motion and stops after passing point cone E (as displayed in Figure 1). The vehicle then follows the same path in reverse (driving backwards) and returns to its start position to complete the test. Without loss of generality, the "lane change to the left" option (as displayed in Figure 1(a)) is used in this paper. The method proposed here can easily be applied to the "lane change to the right" option (as displayed in Figure 1(b)).

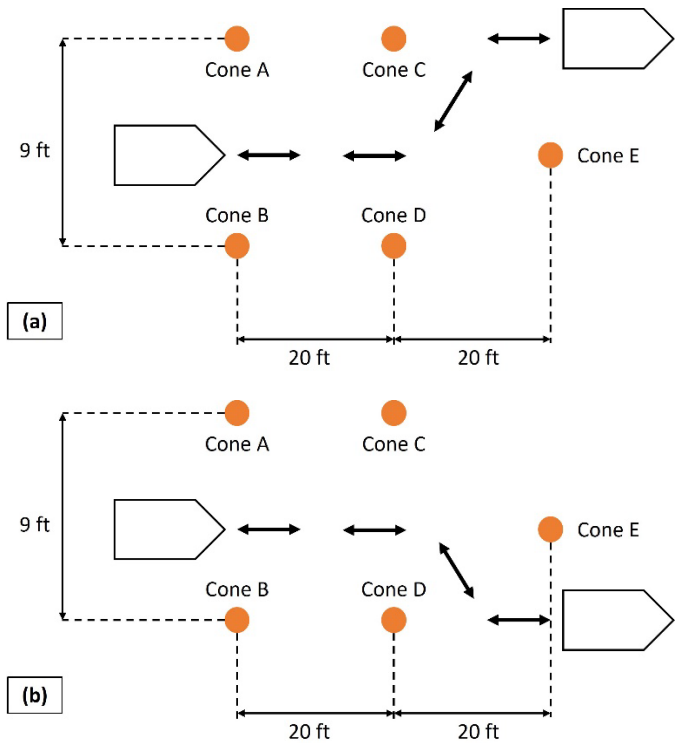

Figure 1. Maneuverability test setup (dimensions from [17]): (a) "lane change to the left" option; (b) "lane change to the right" option

To construct a reference path with smooth curvature such that satisfactory tracking performance can be achieved, the first step is to fix the general shape of the path. Sample waypoints are generated considering the position of the pylons, and a modified Akima piecewise cubic Hermite interpolation routine as discussed in [18], [19] is used to span more waypoints. This procedure is illustrated in Figure 2. The generated waypoints are then divided into segments, where each segment can be represented with a separate polynomial expression. In this case, the waypoints are divided into 4 segments such that each segment includes at most one turning maneuver. The segmented waypoints are displayed in Figure 3.

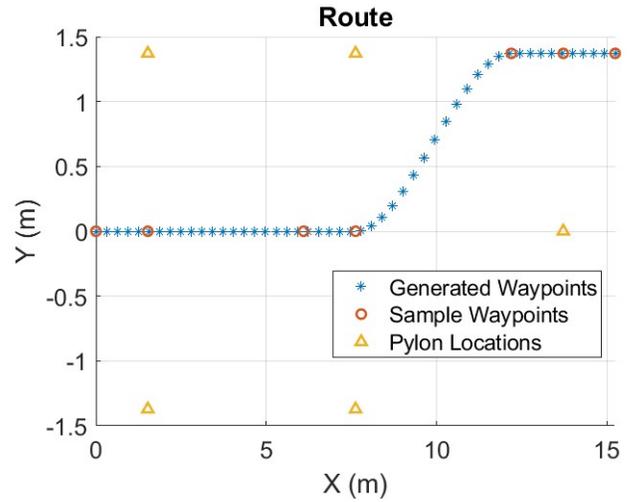

Figure 2. Waypoint generation

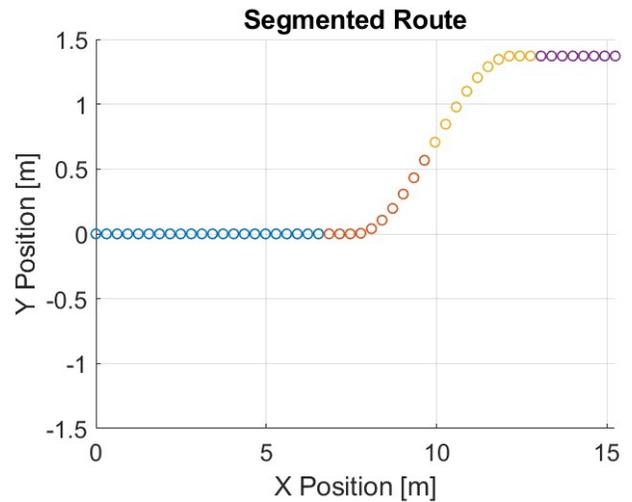

Figure 3. Segmented waypoints

With the path waypoints generated and segmented, polynomial fit optimization procedure can be performed to obtain the desired path expression that guarantees smooth curvature within each segment. The following paragraphs illustrate how this process is set up.

For the ith segment of the path, the polynomial representation of the coordinates can be displayed as follows:

$$\begin{cases} x_i(\lambda) = \sum_{k=0}^{p} a_{i,k} \lambda^k \\ y_i(\lambda) = \sum_{k=0}^{p} b_{i,k} \lambda^k \end{cases} \quad (1)$$

where: $\begin{cases} \lambda \in [0,1] \\ p = polynomial\ order \end{cases}$

The goal is to find the appropriate $a_{i,k}$ and $b_{i,k}$ coefficients for each i and k value. Rearranging Equation 1, one can obtain the regression form of the polynomial. For simplicity, the analysis from this point forward will only be performed on X-coordinates, but the procedure remains the same for Y-coordinates as well. The regression form of X-coordinates of the ith segment can be displayed as follows:

$$x_i = \varphi^T \theta_i \in \mathbb{R} \quad (2)$$

where:
$$\begin{cases} \varphi = [\lambda^0, \lambda^1, \ldots, \lambda^p]^T \in \mathbb{R}^{(p+1) \times 1} \\ \theta_i = [a_{i,0}, a_{i,1}, \ldots, a_{i,p}]^T \in \mathbb{R}^{(p+1) \times 1} \end{cases}$$

In Equation 2 above, $\varphi$ is called the regressor, and the parameters we aim to find out are grouped into the vector $\theta_i$. Assembling the segments together, the overall polynomial regression in matrix form can be displayed as follows:

$$\begin{cases} X = [x_{1,1}, x_{1,2}, \cdots, x_{1,n_1}, \cdots, x_{m,1}, x_{m,2}, \cdots, x_{m,n_m}]^T \in \mathbb{R}^{n \times 1} \\ \Phi = \begin{bmatrix} \varphi^T(\lambda_{1,1}) \\ \varphi^T(\lambda_{1,2}) \\ \vdots \\ \varphi^T(\lambda_{1,n_1}) \\ \vdots \\ \varphi^T(\lambda_{m,1}) \\ \varphi^T(\lambda_{m,2}) \\ \vdots \\ \varphi^T(\lambda_{m,n_m}) \end{bmatrix} \in \mathbb{R}^{n \times (p+1)m} \quad (3) \\ \Theta = [\theta_1^T, \cdots, \theta_m^T]^T \in \mathbb{R}^{(p+1)m \times 1} \end{cases}$$

where: $\begin{cases} m = number\ of\ path\ segment \\ n = n_1 + \cdots + n_m = number\ of\ waypoints \end{cases}$

The optimization problem can hence be written concisely as follows:

$$\min_{\Theta} \|\Phi\Theta - X\|^2 \quad (4)$$

To achieve satisfactory path-tracking control performance, continuous and smooth path curvature is required. To ensure this, continuity constraints in the form of equality constraints are applied to the optimization problem. A continuity constraint of order q between the ith segment and the (i+1)th segment can be written as follows:

$$\begin{cases} x_i(1) = x_{i+1}(0) \\ \frac{dx_i}{d\lambda}(1) = \frac{dx_{i+1}}{d\lambda}(0) \\ \vdots \\ \frac{d^q x_i}{d\lambda^q}(1) = \frac{d^q x_{i+1}}{d\lambda^q}(0) \end{cases} \quad (5)$$

Applying Equation 1 to Equation 5, one can obtain the following:

$$\begin{cases} (a_{i,0} + a_{i,1} + \cdots + a_{i,p}) - a_{i+1,0} = 0 \\ (a_{i,1} + 2a_{i,2} + \cdots + pa_{i,p}) - a_{i+1,1} = 0 \\ \vdots \\ \left(q! a_{i,q} + (q+1)! a_{i,q+1} + \cdots + \frac{p!}{(p-q)!} a_{i,p}\right) - q! a_{i+1,q} = 0 \end{cases} \quad (6)$$

Representing equation 6 in matrix form will yield the following:

$$A_i \theta_i + B_i \theta_{i+1} = 0 \quad (7)$$

where:
$$\begin{cases} A_i = \begin{bmatrix} 1 & 1 & 1 & \cdots & 1 & \cdots & 1 \\ 0 & 1 & 2 & \cdots & q & \cdots & p \\ \vdots & \vdots & \vdots & \ddots & \vdots & \ddots & \vdots \\ 0 & 0 & 0 & \cdots & q! & \cdots & \frac{p!}{(p-q)!} \end{bmatrix} \in \mathbb{R}^{(q+1) \times (p+1)} \\ B_i = \begin{bmatrix} -0! & 0 & \cdots & 0 & \cdots & 0 \\ 0 & -1! & \cdots & 0 & \cdots & 0 \\ \vdots & \vdots & \ddots & 0 & \ddots & 0 \\ 0 & 0 & 0 & -q! & \cdots & 0 \end{bmatrix} \in \mathbb{R}^{(q+1) \times (p+1)} \end{cases}$$

Similarly, continuity constraints between other neighboring segments can be represented in the form of Equation 7 as well. Assuming that the path does not form a closed loop, assembling all the continuity constraint conditions together, one can obtain the following:

$$\Gamma\Theta = 0 \quad (8)$$

where: $\Gamma = \begin{bmatrix} A_1 & B_1 & 0 & \cdots & 0 & 0 \\ 0 & A_2 & B_2 & \cdots & 0 & 0 \\ \vdots & \vdots & \vdots & \ddots & \vdots & \vdots \\ 0 & 0 & 0 & \cdots & A_{m-1} & B_{m-1} \end{bmatrix} \in \mathbb{R}^{(q+1)(m-1) \times (p+1)m}$

The constrained optimization problem can thus be written as follows:

$$\min_{\Theta} \|\Phi\Theta - X\|^2 \quad (9)$$

$$subject\ to: \Gamma\Theta = 0$$

To set up the optimization problem described in Equation 9 for the maneuverability test path, three setting parameters must be specified, namely, $m$, $p$ and $q$. Table 1 displays the value selection for these three setting parameters. As discussed above, the waypoints are divided into 4 segments so that each segment contains at most one turn. With the waypoints segmented as displayed in Figure 3, the polynomial order turns out to yield best curvature result (smallest rms value) at the value of 6. The order of continuity constraint is selected as 3, as order 0 merely guarantees path continuity, order 1 guarantees smooth path at segment transitions, order 2 guarantees path curvature continuity, while order 3 guarantees smooth path curvature at segment transitions.

Table 1. Setting parameters for the segmented polynomial fit optimization of maneuverability test path.

| Setting Parameter | Value | Note |
|---|---|---|
| $m$ | 4 | Number of polynomial segments |
| $p$ | 6 | Polynomial order |
| $q$ | 3 | Order of continuity constraint |

The optimization problem described in Equation 9 can be solved by any standard optimization solver. In this case, the problem is solved by using the MATLAB built-in function fmincon(). Solving the optimization problem for both X and Y-coordinates, the optimized path is shown in Figure 4.

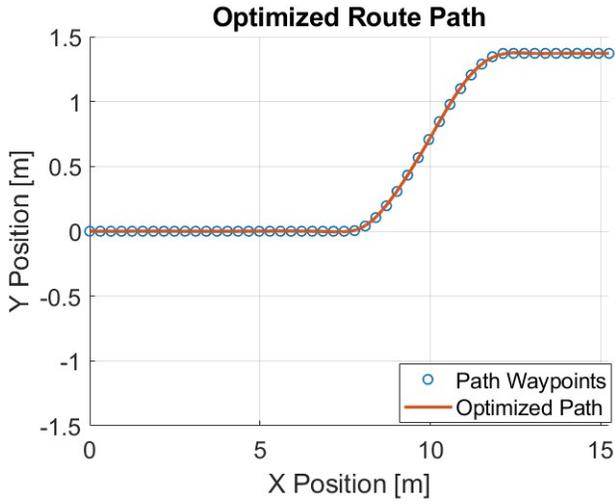

Figure 4. Optimized path

To calculate the path reference curvature, one can apply Equation 10. It should be noted that Equation 10 makes use of second order path derivatives, meaning that only continuity constraints of order 3 and above can guarantee smooth path curvature, as discussed previously.

$$\rho_{ref} = \frac{\frac{dX_p(\lambda)}{d\lambda}\frac{d^2Y_p(\lambda)}{d\lambda^2} - \frac{dY_p(\lambda)}{d\lambda}\frac{d^2X_p(\lambda)}{d\lambda^2}}{((\frac{dX_p(\lambda)}{d\lambda})^2 + (\frac{dY_p(\lambda)}{d\lambda})^2)^{3/2}} \quad (10)$$

where: $\begin{cases} X_p = X \text{ coordinate of optimized path} \\ Y_p = Y \text{ coordinate of optimized path} \end{cases}$

Using the optimized path data and applying Equation 10, the path curvature can be plotted as shown in Figure 5. Since the waypoints generated in Figure 2 are ordered such that the first waypoint is located at the start position of forward motion, the path curvature calculated is for the forward motion as well. To obtain the path reference curvature for backward motion, one can simply flip the forward path curvature and then invert it, the result of which is also displayed in Figure 5.

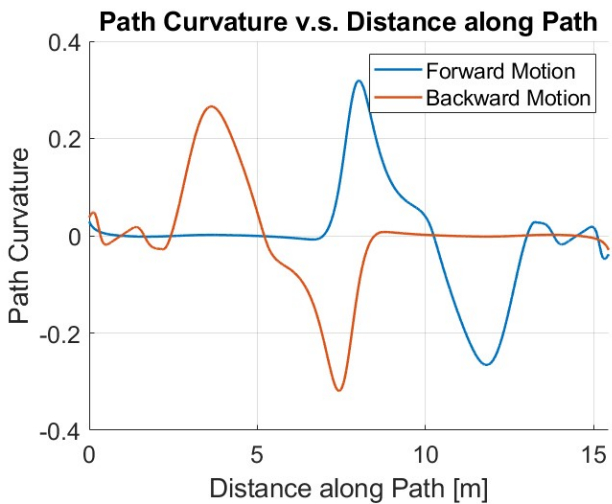

Figure 5. Path curvature for forward and backward motion

## Linear Path-tracking Model

This section aims to present the linear path-tracking model that serves as the basis of control system design. The detailed derivation of this model can be found in [10].

The linear path-tracking error model consists of two components: a (linear) lateral single-track model and a path-tracking model augmentation. The path-tracking scenario is illustrated in Figure 6, and the resultant linear path-tracking model is described in Equation 11. The parameters of the model are listed and described in Table 2.

$$\begin{bmatrix} \dot{\beta} \\ \dot{r} \\ \Delta\dot{\psi}_p \\ \dot{e}_y \end{bmatrix} = \begin{bmatrix} \frac{-C_f-C_r}{MV} & -1+\frac{C_rl_r-C_fl_f}{MV^2} & 0 & 0 \\ \frac{C_rl_r-C_fl_f}{I_z} & \frac{-C_fl_f^2-C_rl_r^2}{I_zV} & 0 & 0 \\ 0 & 1 & 0 & 0 \\ V & l_s & V & 0 \end{bmatrix} \begin{bmatrix} \beta \\ r \\ \Delta\psi_p \\ e_y \end{bmatrix} + \begin{bmatrix} \frac{C_f}{MV} & \frac{C_r}{MV} \\ \frac{C_fl_f}{I_z} & \frac{C_rl_r}{I_z} \\ 0 & 0 \\ 0 & 0 \end{bmatrix} \begin{bmatrix} \delta_f \\ \delta_r \end{bmatrix} + \begin{bmatrix} 0 \\ 0 \\ -V \\ -l_sV \end{bmatrix} \rho_{ref} + \begin{bmatrix} 0 \\ \frac{1}{I_z} \\ 0 \\ 0 \end{bmatrix} M_{zd} \quad (11)$$

where: $l_s = KV$, K is a constant

Table 2. Linear path-tracking model parameters.

| Model Parameter | Explanation |
|---|---|
| $\beta$ | Vehicle side slip angle |
| $r$ | Vehicle yaw rate |
| $\Delta\psi_p$ | Heading error |
| $e_y$ | Path-tracking error |
| $C_f$ | Front tire cornering stiffness |
| $l_f$ | Distance between CG and front axle |
| $C_r$ | Rear tire cornering stiffness |
| $l_r$ | Distance between CG and rear axle |
| $M$ | Vehicle mass |
| $V$ | Vehicle velocity |
| $l_s$ | Preview distance |
| $I_z$ | Vehicle yaw moment of inertia |
| $\rho_{ref}$ | Reference path curvature |
| $M_{zd}$ | Yaw moment disturbance |
| $K$ | Preview distance scheduling constant |

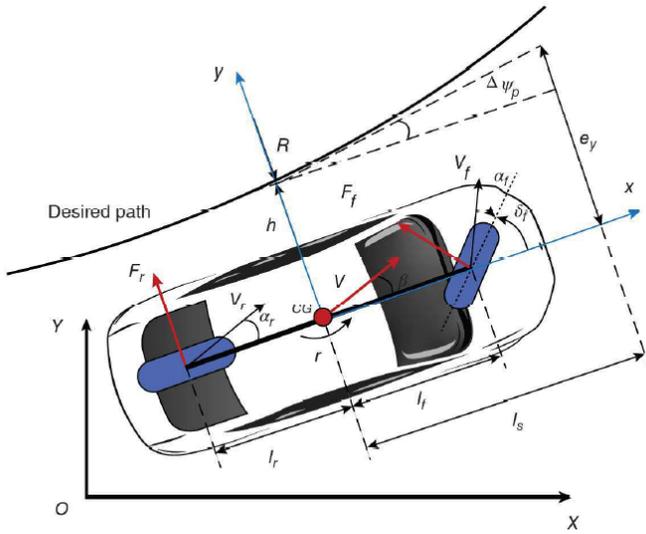

Figure 6. Path-tracking scenario [10]

It can be observed that the model presented in Equation 11 has front and rear steering angle $\delta_f$ and $\delta_r$ as inputs (only front wheel steering shown in Figure 6), and path curvature $\rho_{ref}$ and yaw moment disturbance $M_{zd}$ entering as external disturbances. It should also be noted that this model assumes forward driving. If we assume the vehicle to be front-wheel-steer, reverse driving can then be considered as a rear-wheel-steer vehicle driving forward. Additionally, $C_f, l_f$ and $C_r, l_r$ parameters are to be interchanged to accommodate backward motion. A final note is that the preview distance $l_s$ is chosen to be a linear function of vehicle velocity.

## Speed and Preview Distance Scheduling

As the maneuverability test involves starting from a standstill and coming to a stop, the control system should be able to handle speed variations. Hence, the control system is designed to be speed-scheduled. To do that, a scheduled speed and preview distance profile must be generated. Table 3 lists the conditions chosen for the scheduling procedure, and Figure 7 displays the speed and preview distance profiles obtained. It is worth noting that in general, path-tracking control is more difficult when speed increases, and this is especially true when the vehicle is going in reverse, as the vehicle handling model in reverse motion will become unstable as the speed keeps increasing.

Table 3. Speed & preview distance scheduling conditions.

| Conditions | Value Selection |
|---|---|
| Maximum allowed lateral acceleration | 0.05 g |
| Maximum allowed longitudinal acceleration | 0.05 g |
| Maximum allowed speed | 1 m/sec |
| Minimum allowed speed | 0.1 m/sec |
| Preview distance scheduling constant (K) | 0.5 |

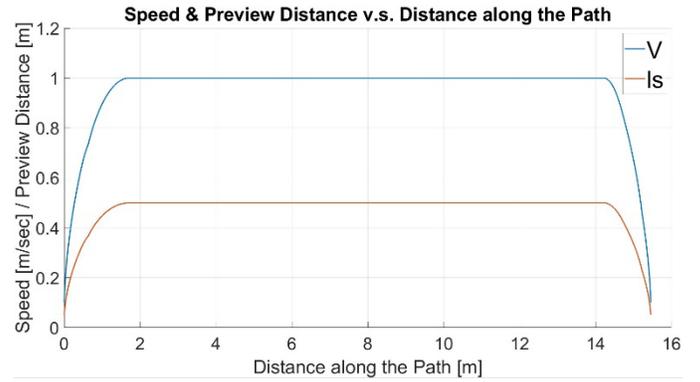

Figure 7. Scheduled speed & preview distance profile

## Disturbance Observer

This section aims to present a general overview of the disturbance observer (DOB). In the application use case of automotive path-tracking, this is also referred to as a curvature rejection filter.

In general, the disturbance observer has two main functions: disturbance rejection and model regulation. To see these, one can first consider a simple input-output system consisting of a plant ($G$) with multiplicative model uncertainty ($\Delta_m$) and an external disturbance ($d$) applied at the output, as displayed in Figure 8. It should be noted that $G_n$ is called the nominal plant. One can, thus, write Equation 12 to describe this system, and further obtain Equation 13 naturally.

$$y = Gu + d = (G_n(1 + \Delta_m))u + d = G_n u + (G_n \Delta_m u + d) = G_n u + e \quad (12)$$

$$e = y - G_n u \quad (13)$$

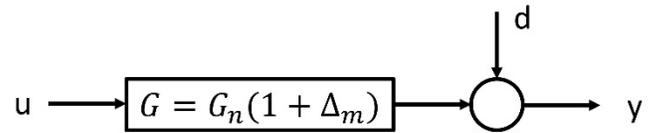

Figure 8. Sample input-output system with disturbance and model uncertainty

Since the purpose of the DOB is to achieve disturbance rejection and model regulation as mentioned above, the end goal is to have a system without model uncertainty and external disturbance, which can be described in Equation 14.

$$y = G_n u_n \quad (14)$$

To determine the input (u) that is necessary to achieve the desired result in Equation 14, one can combine Equation 12 and Equation 14, and further apply Equation 13 to get the result described in equation 15.

$$u = u_n - \frac{y}{G_n} + u \quad (15)$$

It should be noted that the result in Equation 15 cannot be implemented for the following two reasons: 1) The input (u) is on both sides of the expression; 2) $1/G_n$ is not proper since $G_n$ is proper for most physical systems. In order to make this result implementable, Equation 15 is modified to Equation 16, where Q is a unity gain low-pass filter of the appropriate order such that $Q/G_n$ is proper.

$$u = u_n - \frac{Q}{G_n}y + Qu \quad (16)$$

Equation 16 can be represented as the system block diagram shown in Figure 9. It should be noted that an additional component (*n*) is added as sensor noise. Using this block diagram, several transfer functions can be derived to demonstrate the functions of DOB, as shown in Equation 17.

$$\begin{cases} \frac{y}{u_n} = \frac{G_n G}{G_n(1-Q)+GQ} \\ \frac{y}{d} = \frac{G_n(1-Q)}{G_n(1-Q)+GQ} \\ \frac{y}{n} = \frac{-GQ}{G_n(1-Q)+GQ} \end{cases} \quad (17)$$

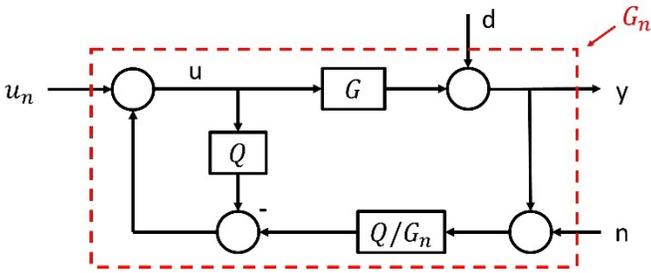

Figure 9. DOB block diagram

From equation 17, it can be observed that at low frequencies when $Q = 1$, $y/u_n = G_n$ and $y/d = 0$, indicating that model regulation and disturbance rejection are both achieved. Additionally, at high frequencies when $Q = 0$, $y/n = 0$, meaning that high frequency sensor noises can be rejected. It can, thus, be concluded that this system structure can achieve the desired effect.

In general, to design a DOB, one should construct a unity gain low-pass filter $Q$ and a nominal plant $G_n$. As discussed in a previous paragraph, the $Q$ filter must be of appropriate order such that $Q/G_n$ is proper. Additionally, the $Q$ filter should also have an appropriate bandwidth to ensure its performance, as a bandwidth too low will result in poor disturbance rejection and model regulation, while a bandwidth too high will cause high frequency sensor noises to enter the system and will also cause stability robustness problems. The nominal plant $G_n$ should also be designed with care, since its dynamics must not be too drastically different from that of the uncertain plant $G$, or else the design will fail to achieve its goals.

Several remarks should be made for the application of DOB on maneuverability test path-tracking control. The plant model $G$ in this case can be represented as the transfer function from front steering angle $\delta_f$ to path-tracking error $e_y$, as we assume the vehicle is front-wheel-steer only. The external disturbance entering the system is the reference path curvature as yaw moment disturbance is assumed to be zero. As a result, $u_n$ and $y$ as displayed in Figure 9 are front steering angle $\delta_f$ and path-tracking error $e_y$, respectively. Since the main goal of this control system is to achieve satisfactory tracking performance, model regulation is not of any particular importance. Hence, nominal plant $G_n$ is simply chosen as $(1.01 \cdot G)$. From the linear path-tracking model in Equation 11, one can derive that plant G is proper and has a relative order of 2, which means that the $Q$ filter must be at least second order to guarantee that $Q/G_n$ is proper. Thus, the $Q$ filter is chosen as a standard second order system with unity gain in the form displayed in Equation 18. Additionally, it is worth noting that since the control system is speed-scheduled and the path-tracking model in Equation 11 includes vehicle velocity and preview distance, the system parameters will vary as the vehicle moves along the path.

$$Q = \frac{\omega_n^2}{s^2 + 2\xi\omega_n s + \omega_n^2} \quad (18)$$

One final remark is that the DOB is usually applied together with an additional feedback controller to further improve control system performance. Figure 10 displays the system block diagram under this setup, where C denotes the feedback controller. The controller C is then designed for the desired or nominal plant model $G_n$.

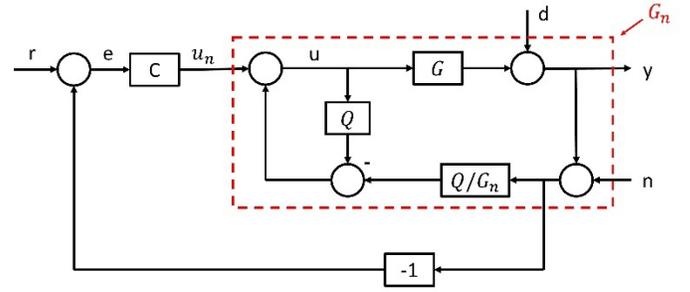

Figure 10. Control system with DOB and feedback controller

## Speed-Scheduled, Parameter-Space Robust PID Controller

As shown in Figure 10, a feedback controller is typically used together with DOB to improve control system tracking performance. In this case, a speed-scheduled, parameter-space PID controller is constructed as the feedback controller to augment the DOB. The parameter-space method is discussed in detail in [15].

Since the plant of this control system is the transfer function from front steering angle $\delta_f$ to path-tracking error $e_y$, the reference input (denoted as *r* in Figure 10) should be set as zero, as we intend to minimize the tracking error. The PID controller form is displayed in Equation 19. The controller gains $(k_p, k_i, k_d)$ are the parameters to be tuned. Considering that the controller is speed-scheduled as well, the tunable parameter set has four elements: $(V, k_p, k_i, k_d)$.

$$C(s) = k_p + \frac{k_i}{s} + k_d s \quad (19)$$

A D-stability region is selected for the parameter-space design for placement of the controlled system poles and is shown in Figure 11. Part of the D-stability based controller parameter space results for forward motion (at maximum speed and minimum speed) are displayed in Figure 12 and Figure 13. When it comes to the value selection of the controller gains, a general rule of thumb is to choose the gains to be as small as possible within the admissible region so that the control efforts can be minimized while the energy efficiency can be maximized. In this case, the controller gain values are selected on the d1 CRB boundary.

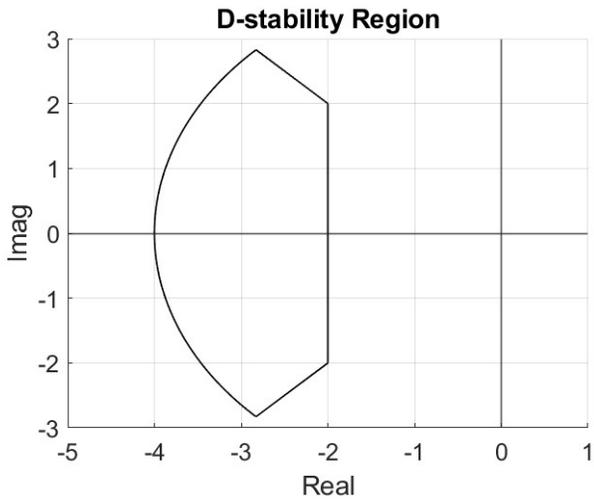

Figure 11. D-stability region

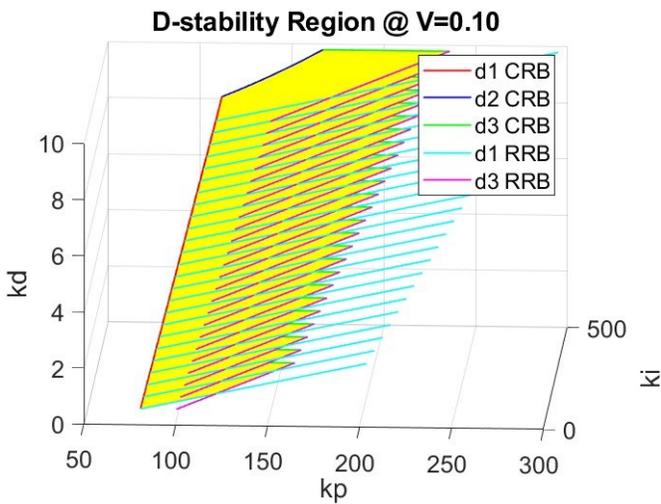

Figure 12. Admissible control region for forward motion at minimum speed

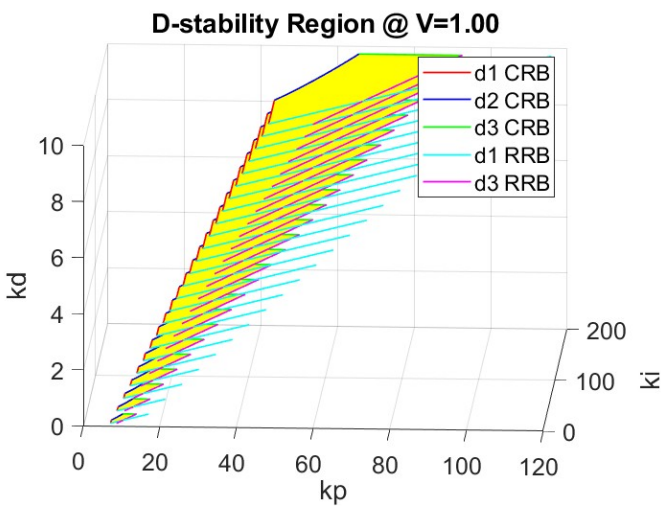

Figure 13. Admissible control region for backward motion at maximum speed

## Simulation Study

Simulation studies are conducted to demonstrate the efficacy of the proposed control system design. A Simulink model is constructed to simulate the motions of the vehicle. The parameter values used in the simulations are listed in Table 4.

Table 4. Parameter value selections for simulation.

| Parameter | Value Selection |
| --- | --- |
| $C_f$ [N/rad] | 3e5 |
| $l_f$ [m] | 2 |
| $C_r$ [N/rad] | 3e5 |
| $l_r$ [m] | 2 |
| M [kg] | 3000 |
| $I_z$ [$kg \cdot m^2$] | 5.113e3 |
| $\omega_n$ [rad/sec] | 100 |
| $\xi$ [unitless] | 0.707 |

For forward motion, the initial state of the vehicle is assigned as shown in equation 20. The simulation results for standalone DOB, standalone PID and combined PID and DOB control system are displayed in Figure 14, Figure 15 and Figure 16, respectively.

$$(X_{initial}, Y_{initial}, \psi_{initial}) = (0,0,0) \quad (20)$$

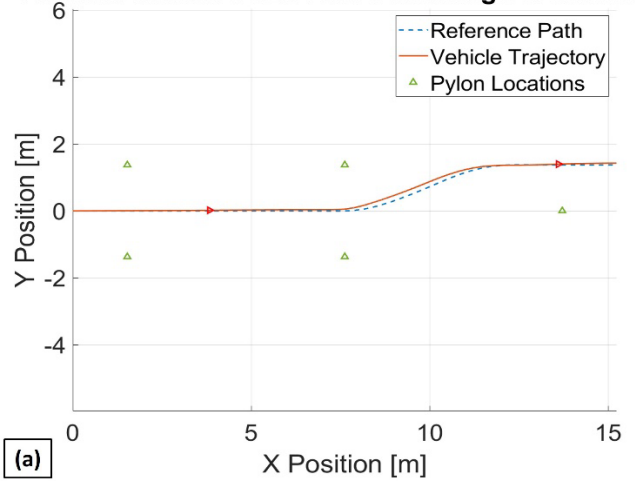

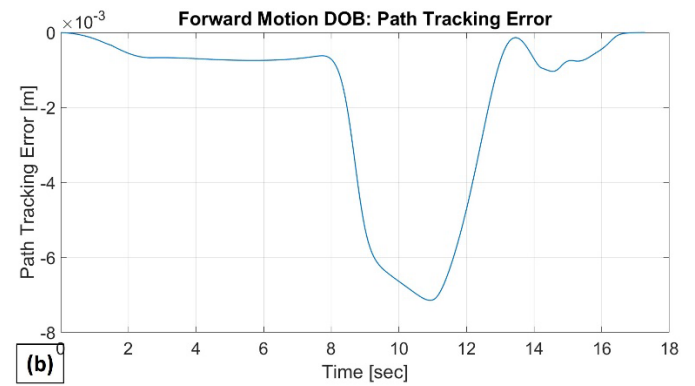

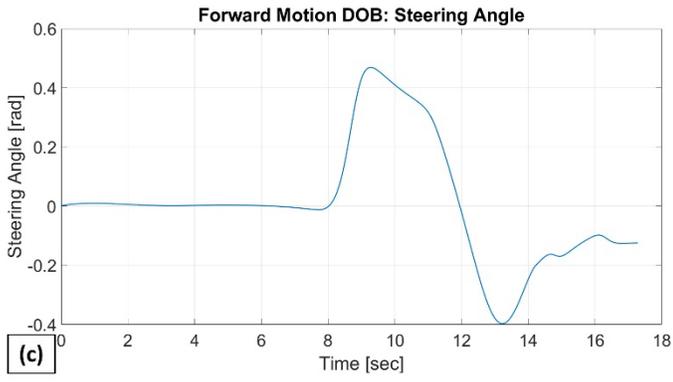
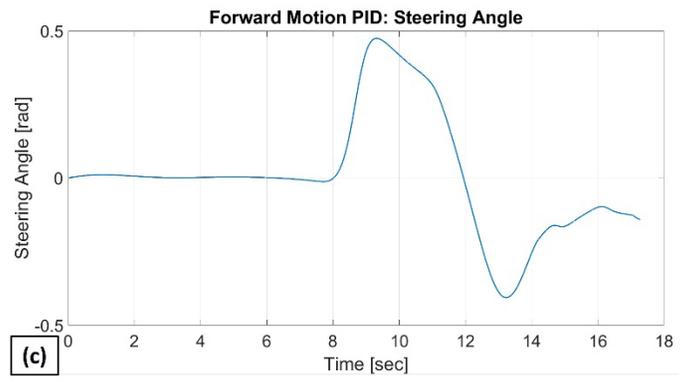

Figure 14. Forward motion DOB simulation results

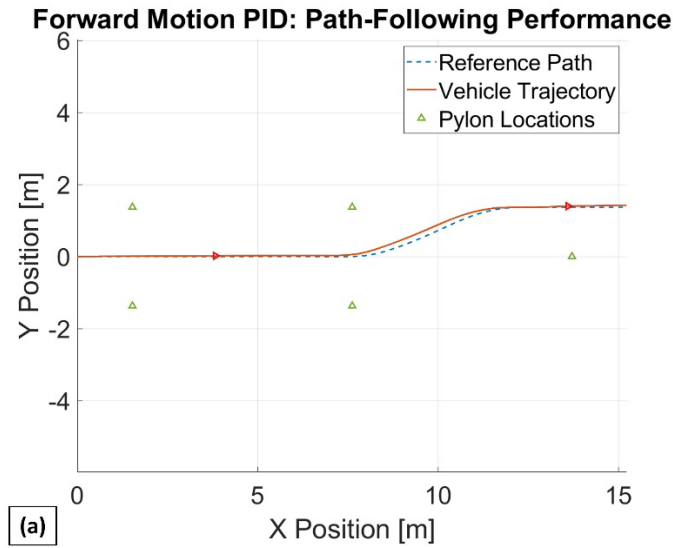
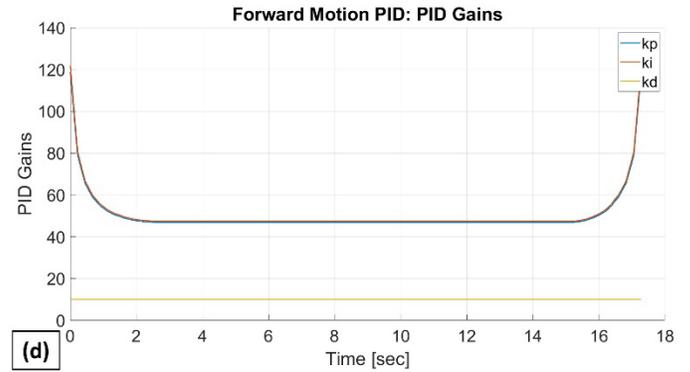

Figure 15. Forward motion PID simulation results

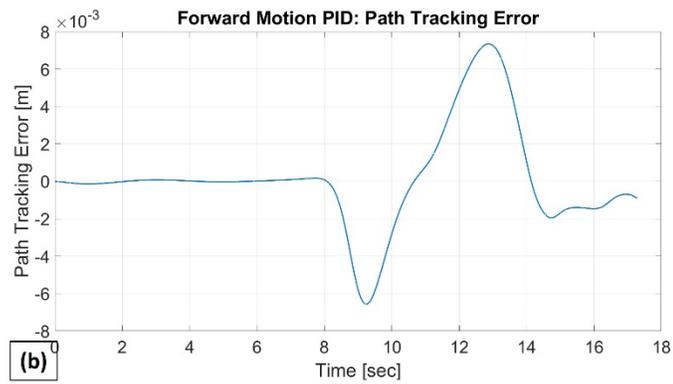
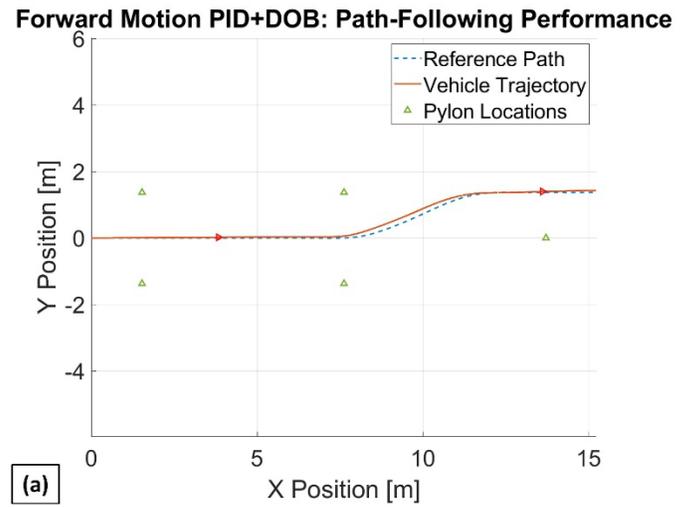
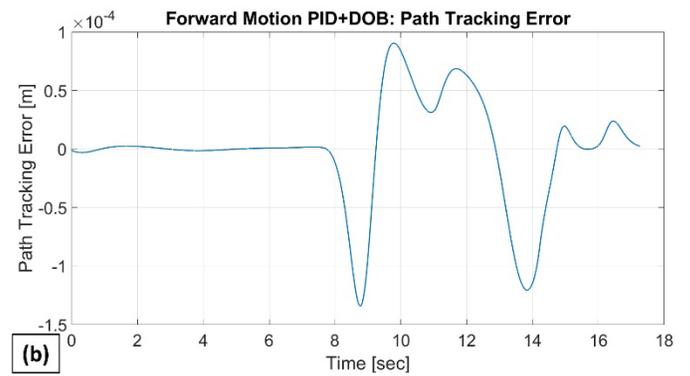

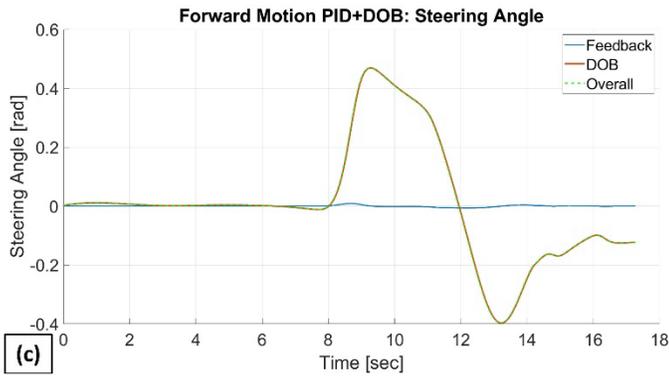

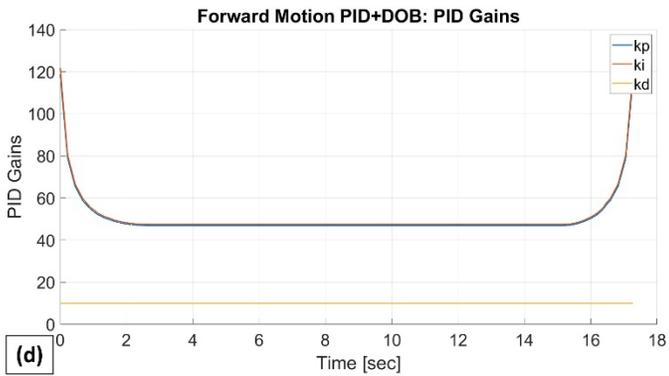

Figure 16. Forward motion PID+DOB simulation results

It can be observed from Figures 14-16 that all three simulations display satisfactory path-tracking performance. To better compare the results, Table 5 is constructed to record maximum absolute path-tracking error, RMS path-tracking error, maximum absolute steering angle as well as maximum absolute steering rate.

Table 5. Forward motion simulation results evaluation.

| Parameter | DOB | PID | PID+DOB |
|---|---|---|---|
| Max absolute path-tracking error [m] | 0.0071 | 0.0073 | 1.3399e-4 |
| RMS path-tracking error [m] | 0.0029 | 0.0027 | 4.4357e-5 |
| Max absolute steering angle [rad] | 0.4692 | 0.4746 | 0.4697 |
| Max absolute steering rate [rad/sec] | 0.6597 | 2.4348 | 2.4538 |

From Table 5, it can be observed that the control system that combines the PID and DOB shows the best performance in terms of minimizing path-tracking errors. It should also be noted from Figure 16 that with the combined DOB and PID control system, the DOB provides the majority of the control action, while the PID controller serves mainly to smooth out the steering input generated by the DOB.

Similar simulation procedure is carried out for the backward motion as well. For backward motion, the initial state of the vehicle can be assigned as shown in equation 21. The simulation results for standalone DOB, standalone PID and combined PID and DOB control system are displayed in Figure 17, Figure 18 and Figure 19, respectively.

$$(X_{initial}, Y_{initial}, \psi_{initial}) = (15.2386, 1.3716, \pi) \qquad (21)$$

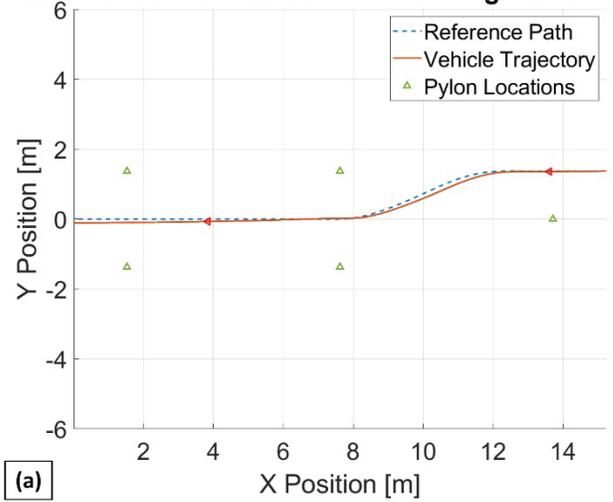

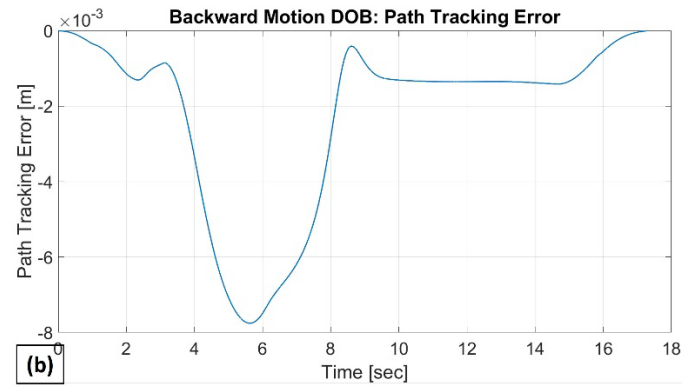

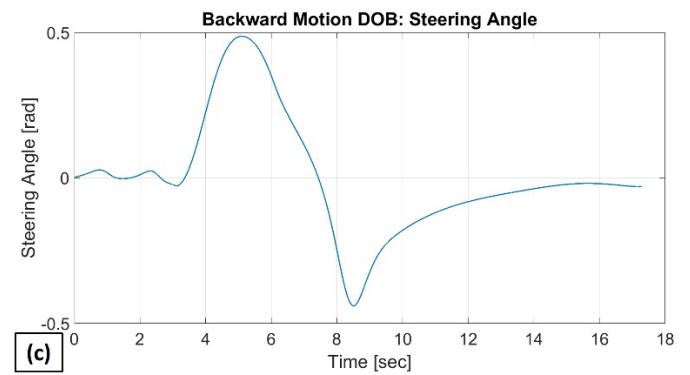

Figure 17. Backward motion DOB simulation results

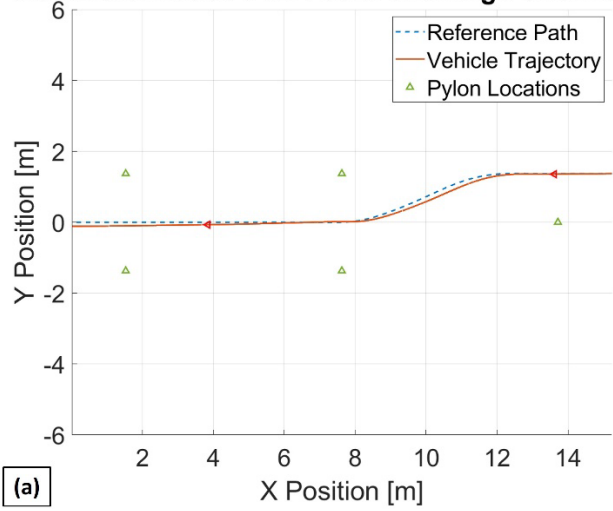
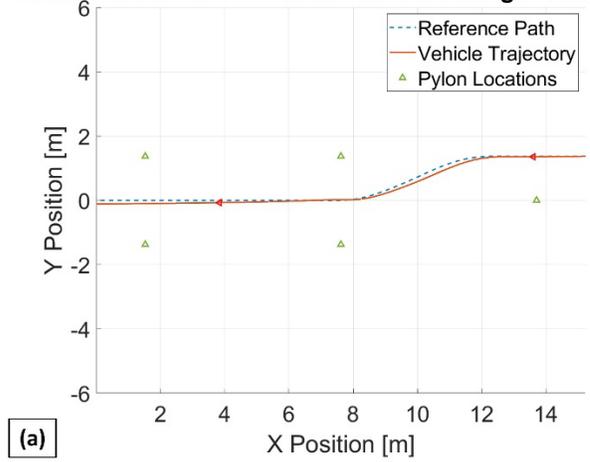
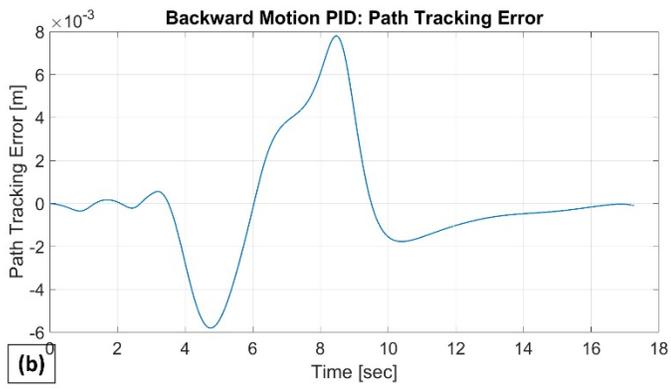
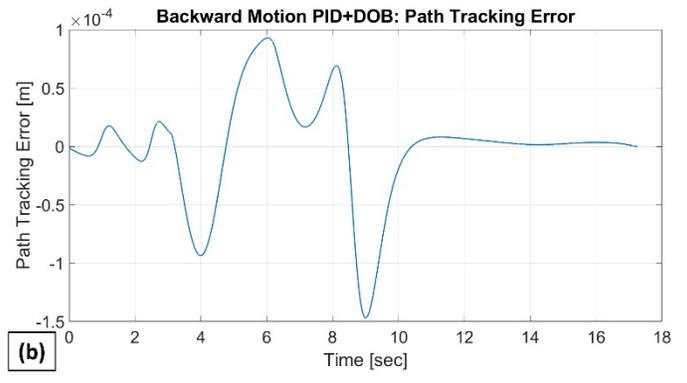
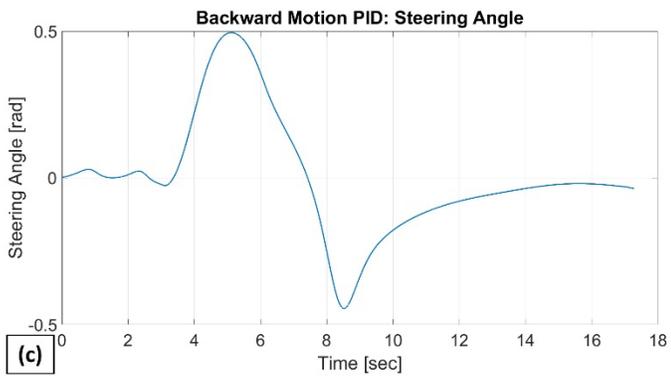
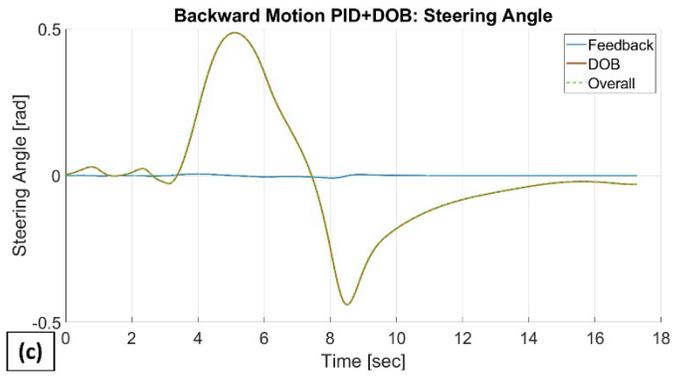
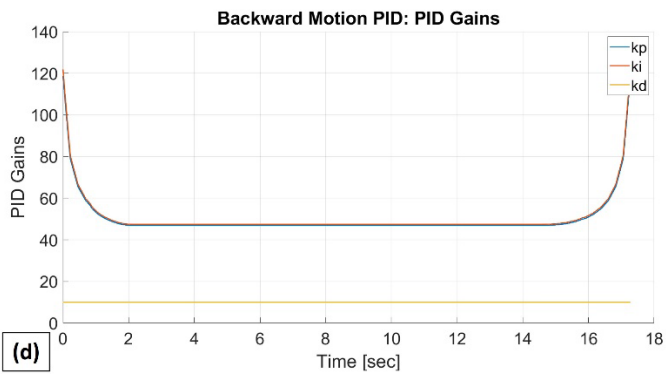
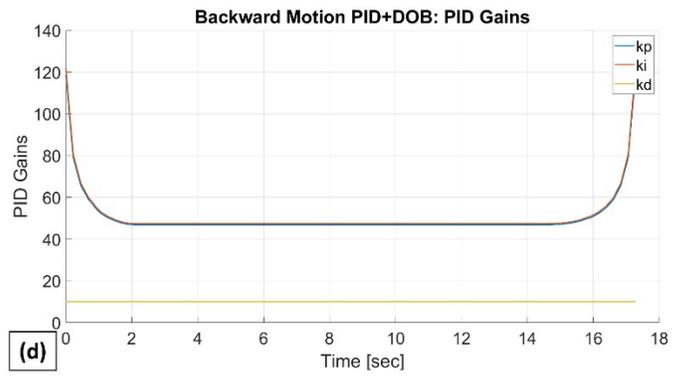

Figure 18. Backward motion PID simulation results

Figure 19. Backward motion PID+DOB simulation results

Similar to forward motion, it can be observed from Figures 17-19 that the tracking performance in reverse for all three cases are satisfactory. Hence, Table 6 can be constructed in a similar fashion as Table 5 to analyze the simulation results.

Table 6. Backward motion simulation results evaluation.

| Parameter | DOB | PID | PID+DOB |
|---|---|---|---|
| Max absolute path-tracking error [m] | 0.0078 | 0.0078 | 1.4684e-4 |
| RMS path-tracking error [m] | 0.0032 | 0.0027 | 4.4335e-5 |
| Max absolute steering angle [rad] | 0.4876 | 0.4951 | 0.4876 |
| Max absolute steering rate [rad/sec] | 0.5561 | 3.16 | 3.1847 |

It can again be observed that in terms of path-tracking error, the combined control system consisting of PID and DOB yields the best results. Again, the DOB generates most of the control efforts, while PID smooths out the steering input. One additional remark is that for both forward and backward motion, the combination of PID and DOB seems to generate the highest steering rate. While the steering rates achieved in the simulation are manageable, one should pay more attention to this metric when conducting tests.

## Summary and Future Work

This paper aims to automate the maneuverability test, which is part of Ohio's driver license test. The path generation problem is set up as a segmented polynomial fit optimization problem with continuity constraints to guarantee path curvature smoothness. A linear path-tracking model is used as the basis for the control system design, which consisted of a disturbance observer (curvature rejection filter) and a speed-scheduled, parameter-space robust PID controller. Simulation studies were conducted, and results indicate that the combination of PID and DOB will yield the best tracking performance.

Future work can use higher fidelity models in the longitudinal [20], lateral [21] and vertical directions [22] or a canned program like Carsim [23], [24]. Experimental work using the VVE method can also be used for future work [25], [26]. The VVE method can be used to park between cars in a virtual environment like an urban one in a smart city [27], [28] for example and for collision avoidance studies [29], [30] also along with path planning and path tracking and even platooning studies [31] with virtual other vehicles in a platoon for coordinated path planning and tracking of the platoon.

## Contact Information

Xincheng Cao: cao.1375@osu.edu

## Acknowledgments

The authors thank the Automated Driving Lab at the Ohio State University for its support. The authors thank NVIDIA for its GPU donations.